# Normal zone propagation along superconducting films deposited on wide substrates

V. Sokolovsky, V. Meerovich[*]

Physics Depatment, Ben-Gurion University of the Negev, P. O. Box 653, Beer-Sheva, 84105, Israel

**Abstract**

The propagation of a normal zone in a long superconducting film on a dielectric or conducting substrate is analyzed analytically. The obtained expressions for the propagation velocity allow one to analyze the effect of the thermal resistance between the film and substrate and the width of the substrate. The results are expended into the cases of the normal zone propagation in a meander and in parallel films and can be approximately applied to multifilamentary coated superconductors.

1. **Introduction**

Structures consisting of layers of high–temperature and $MgB_2$ superconductors deposited on conducting and dielectric substrates are very promising for use in electronics and power electric applications [1-4]. In recent years, the prospective applications of these structures have widened with the development of coated conductors [3]. One of the most important problems to be solved for implementation of superconductors is their thermal stability. The thermal state of a film or a coated superconductor is determined by the competition between losses and heat removal where the substrate plays a substantial role. To investigate the thermal stability, two types of problems are usually considered. The first type is devoted to the study of consequences caused by a thermal disturbance due to energy release in an infinite line source along a superconductor [5]. The second type of problems is related to the normal zone propagation along a superconductor [6-8]. These problems are important for both the protection of the superconducting devices from the quenching and providing the fast normal zone propagation in current limiters and switches [1,4,8-10]. To solve these problems, it is frequently assumed that the temperature of the film and substrate changes only along the film and does not vary in the direction across the structure. The analytical approach developed in [11] for a superconducting film deposited on a wide substrate has shown that account of temperature variation in the direction across the structure leaded to new effects even for the case of uniform temperature distribution along the film.

[*] Corresponding author: Tel.: +972 8 647 2458; fax: +972 8 647 2903.
E-mail address: victorm@bgu.ac.il (V. Meerovich).

Two- and three-dimensional models for the formation and propagation of normal zones in the superconducting structures are usually based on numerical approach [12-14]. However, it is difficult from the obtained numerical results to determine practically important parameters such as the minimal normal-zone propagating current and the propagation velocity.

In this paper we present analytical expressions describing the velocity of the normal zone propagation in a film deposited on a wide substrate.

In section 2 we develop a mathematical model for a 2D-structure of the narrow superconducting film deposited on a wide dielectric substrate. The structure models various superconducting elements of electronic and power devices, such as fast operating switches, current limiters [1,4,9,10,14], as well as multifilament coated superconductors where an electrical contact between superconducting strips and a conducting substrate is practically absent [15]. Analytical solutions for several practical cases are obtained in section 3. In section 4 the model is expanded into the structures where a current is distributed between a superconducting film and a conducting substrate. This case describes the multifilament coated superconductors with an electrical contact between a superconductor and a substrate [16] or the film structures where a superconductor is deposited on a conducting substrate without a buffer layer [17,18].

## 2. Mathematical Model

Let us consider a thin structure presented in Fig. 1. The specific heat fluxes from a superconducting film, $Q_1$, $Q_2$, and from a substrate, $Q_1$, $Q_3$, are shown in this figure. In the general case, the heat transfers from various sides of the substrate are different: $Q_3 \neq Q_1$. Note that $Q_2 = 0$ at $|y| \geq L$, where $2L$ is the width of the film. We assume that all the heat fluxes are determined by the Newton law: they are proportional to the difference of the temperatures:

$$Q_1 = H_1 (T_f - T_0),\ Q_2 = H_2 (T_f - T_s),\ \text{and}\ Q_3 = H_3 (T_s - T_0), \qquad (1)$$

where indexes $f$ and $s$ note the parameters of the film and substrate, respectively; $T$ is the temperature; $T_0$ is the temperature of the coolant; $H_i$ ($i = 1, 2, 3$) are the heat transfer coefficients. The value of the heat transfer coefficient $H_2$ characterizes the thermal resistance of the contact between the superconducting film and substrate. In the limiting cases: $H_2 \to 0$ and $H_2 \to \infty$ the problem is much simplified. In the first case one can neglect the heat transfer from the film to the substrate and the velocity of the normal zone propagation is described by the expression obtained for superconducting wires [19-21]. In another limiting case $H_2 \to \infty$, temperatures of the film and substrate are the same at $|y| \leq L$.

Assuming that the specific heat capacity $C$ and thermal conductivity $\lambda$ are temperature-independent, the thermal state of the structure can be described by the following dimensionless equations [11]:

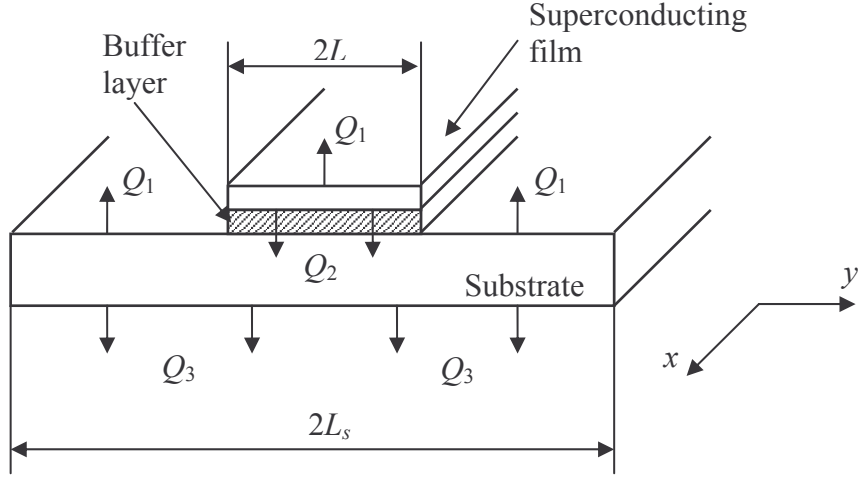

Fig. 1. Sketch of a SC film on a wide substrate.

$$\alpha u j - h_1 \tau_f - h_2(\tau_f - \tau_s) = 0, \qquad (2)$$

$$\frac{\partial \tau_s}{\partial t^*} = \frac{\partial^2 \tau_s}{\partial x^{*2}} + \frac{\partial^2 \tau_s}{\partial y^{*2}} - h_3 \tau_s + h_2(\tau_f - \tau_s), \qquad \text{at } |y^*| \le l \qquad (3)$$

$$\frac{\partial \tau_s}{\partial t^*} = \frac{\partial^2 \tau_s}{\partial x^{*2}} + \frac{\partial^2 \tau_s}{\partial y^{*2}} - h \tau_s, \qquad \text{at } |y^*| > l \qquad (4)$$

$$2li = \int_{-l}^{l} j \, d\xi \qquad (5)$$

where $j = J/J_c(T_0)$; $\tau_j = (T_j - T_0)/(T_c - T_0)$ at $j = s, f$; $J$ is the local current density; $t^* = tH_z/C_s(T_0)\Delta_s$; $J_c$ is the critical current density; $u = U/\rho(T_0)J_c(T_0)$; $U$ is the voltage drop per unit of the film length; $x^* = x/L_x$; $y^* = y/L_x$; $l = L/L_x$; $h_k = H_k(T)/H_z$ at $k = 1,2,3$; $h = [H_1(T)+H_3(T)]/H_z$; $H_z = H_1(T_0)+H_3(T_0)$; $\alpha = \rho(T_0)J_c(T_0)^2 \Delta_f/[H_z(T_c-T_0)]$ is the analog of the so called Stekly's parameter; $i = I/I_c(T_0)$; $\lambda_s$ is the thermal conductivity of the substrate; $\Delta_s$ is the thickness of the substrate; $I$ is the current in the film; $I_c(T_0) = 2LJ_c(T_0)\Delta_f$ is the critical current of the film at $T = T_0$; $L_x = [\lambda_s(T_0)\Delta_s/H_z]^{1/2}$ and is about 1 cm for a sapphire substrate at 77 K [11]. Henceforth, the symbol "*" is omitted in the notations of the dimensionless values.

The film temperature is determined by the losses $\alpha u i$ and two heat fluxes: to the coolant $h_1 \tau_f$ and to the substrate $h_2(\tau_f - \tau_s)$. At $|y| < l$ the temperature of the substrate is determined by the last flux, the heat flux power to the coolant, $\tau_s h_3$, and the heat flux power through the substrate.

The boundary conditions for Eqs. (3), (4) are determined by the continuity of the temperature and heat fluxes at the film boundaries and by neglecting the heat flux from face planes of the substrate at $y = \pm l_s$ (Fig. 1), which is proportional to the thickness $\Delta_s \leq 1$ mm.

Since we consider a narrow film, $l \ll 1$, the temperature across the film can be taken constant so that the current density $j$ is independent of $y$. Integration of Eq. (5) gives $j = i$. Then integrating Eq. (3) with taking into account Eq. (2) and keeping only terms which do no contain $l$, we obtain

$$\left.\frac{\partial \tau_s}{\partial y}\right|_{y=l} - \left.\frac{\partial \tau_s}{\partial y}\right|_{y=-l} + \alpha_{ef} u i = 0, \tag{6}$$

where $\alpha_{ef} = 2l h_2 \alpha / (h_1 + h_2)$. Note that the value $2l\alpha u i$ is finite and characterizes the total losses per unit of the film length.

The voltage drop across a superconductor depends on the local current density and temperature. To investigate the thermal state of a film with nonuniform distribution of the temperature along a narrow film ($l \ll 1$), we use the stepwise voltage-current characteristic:

$$u = \begin{cases} 0, & i < 1 - \tau_f, \\ i, & i \geq 1 - \tau_f. \end{cases} \tag{7}$$

The critical current is a linear function of the temperature $I_c(T) = I_c(T_0)(1-\tau_f)$, and the film resistance jumps from zero up to 1 at the critical current. This approximation is frequently used to obtain analytical expressions for the analysis of the normal zone propagation and thermal domains [19-21]. The temperature of the film in Eq. (7) is determined from (2):

$$\tau_f = \frac{\alpha i^2}{h_1 + h_2} + \frac{h_2 \tau_s}{h_1 + h_2}. \tag{8}$$

Here we assume that all $h_i$ are independent of temperature. If $v$ is the steady-state velocity of a normal zone, Eq. (4) can be rewritten as ($h = 1$):

$$-v\frac{\partial \tau_s}{\partial X} = \frac{\partial^2 \tau_s}{\partial^2 X} + \frac{\partial^2 \tau_s}{\partial^2 y} - \tau_s, \tag{9}$$

where $X = x - vt$ and the normal zone occupies the film at $X < 0$.

In the chosen dimensionless units the characteristic velocity is

$$v_x = \frac{1}{C}\sqrt{\frac{\lambda_s H_z}{\Delta_s}}$$

and this velocity is estimated as about 0.4 m/s for sapphire substrate of 1 mm thickness (for sapphire $\lambda_s \approx 1100$ W/m·K and $C \approx 2.4 \times 10^5$ J/m$^3$K at 77 K [22,8] and $H_z \approx 10^4$ W/m$^2$K). The characteristic velocity in high temperature superconductors with the same thickness is about 100 times less due to their very low heat conductivity (~1 W/m·K) and higher heat capacity ($C \approx 7.5 \times 10^5$ J/m$^3$K) [23].

### 3. Normal zone propagation in films on a dielectric substrate

*3.1. A single film on an infinitely wide substrate*

Let us first consider the propagation of the normal zone in a single straight film of an infinite length deposited on an infinitely wide substrate ($l_s \gg 1$). Due to symmetry of the task it is enough to consider a half of the substrate at $y > 0$. The boundary conditions for Eq. (9) are:

$\tau_s = 0$ at $y \to \infty$ or $X \to \infty$;

$\dfrac{\partial \tau_s}{\partial y} = -\alpha_{ef} i^2 / 2$ at $y = 0$ and $X \leq 0$;

$\dfrac{\partial \tau_s}{\partial y} = 0$ at $y = 0$ and $X > 0$.

The temperature at $X \to -\infty$ is given by a solution of Eq. (9) where the derivatives with respect to $X$ equal to zero:

$$\tau_s = \frac{\alpha_{ef} i^2}{2} e^{-y}. \tag{10}$$

The velocity $v$ is determined from the condition that the current equals to the critical value at the boundary between the normal and superconducting zones:

$$i = 1-\tau_f \qquad \text{at } X = 0 \text{ and } y = 0. \tag{11}$$

The task is reduced to the Helmholtz equation:

$$\frac{\partial^2 w}{\partial X^2} + \frac{\partial^2 w}{\partial y^2} = \left(1 + \frac{v^2}{4}\right) w, \tag{12}$$

where $\tau_s = w \exp(-vX/2)$.

Using the solution for Eq. (12) presented in [24], we obtain the solution of Eq. (9) in the following form:

$$\tau_s(X,y) = \frac{\alpha_{ef} i^2}{2} FT(X,y,v), \tag{13}$$

where $FT(X,y,v) = \frac{1}{\pi} \int_{-\infty}^{0} e^{\frac{v(\xi-X)}{2}} K_0\left(\sqrt{\left(1+\frac{v^2}{4}\right)\left[(X-\xi)^2 + y^2\right]}\right) d\xi$, $K_0$ is the modified Bessel function of the zero order. Note that at $x \to -\infty$ $K_0(x) \approx e^{-x}/\sqrt{2x/\pi}$ and the integral in (13) has a finite magnitude at any $v$.

Taking into account of (8), (11) and (13), the task of the velocity determination is reduced to a solution of the following equation:

$$1 - i = \alpha i^2 \left[\gamma + \frac{\beta}{\pi} FT(0,0,v)\right], \tag{14}$$

where $\beta = h_2^2 l/(h_1+h_2)^2$ and $\gamma = 1/(h_1+h_2)$.

The velocity is determined by not only the Stekly parameter $\alpha$ as for a wire, but also by the parameters $\beta$ and $\gamma$ characterizing the thermal resistance between a film and a substrate. The velocity $v$ as a function of the current $i$ is presented in Fig. 2 for various $\gamma$ and $\beta$.

One important parameter is the minimal current $i_m$ of the zone propagation [19-21], which is corresponding to $v = 0$. In this case the integral in Eq. (13) gives $\pi/2$ and this current is

$$i_m = \frac{\sqrt{1+2\alpha_{im}}-1}{\alpha_{im}}, \qquad (15)$$

where $\alpha_{im} = 2\alpha(\gamma + \beta/2)$.

Expression (15) is congruent to the expression for the minimal current of the normal zone propagation in a superconducting wire [19-21], where the Stekly parameter is replaced by a parameter $\alpha_{im}$.

The parameters $h_i$ characterize the significance of various heat fluxes from the superconducting film to the coolant. If the cooling intensities from two sides of the structure are the same, for example, the structure is situated vertically, then $h_3 = h_1$. In the opposite case, where the cooling is realized only from one side, one of these parameters is zero. The heat transfer coefficient $h_2$ depends on the technology of the film deposition as well as on the material of the buffer layer and its thickness, which is usually in the range $10 \div 100$ nm. At $h_2 \gg 1$ the task is simplified: the temperatures of the film and substrate are the same. Parameters $\beta$ and $\gamma$ in Eq. (14) become $\beta = l$ and $\gamma = 0$. The cooling of the film is realized due to the heat flux through the substrate and for the symmetrical task the flux is the same in both directions:

$$\left|\frac{\partial \tau_s}{\partial y}\right|_{y=0_{-0}} = \left|\frac{\partial \tau_s}{\partial y}\right|_{y=0_{+0}} = \alpha l i^2.$$

It is convenient to introduce here the effective Stekly parameter as $\alpha'_{ef} = l\alpha$. Then the minimal current $i_m$ of the zone propagation is given by expression (15) where $\alpha_{im}$ is replaced by $\alpha'_{ef}$. Fig. 3 gives the temperature distribution for this case.

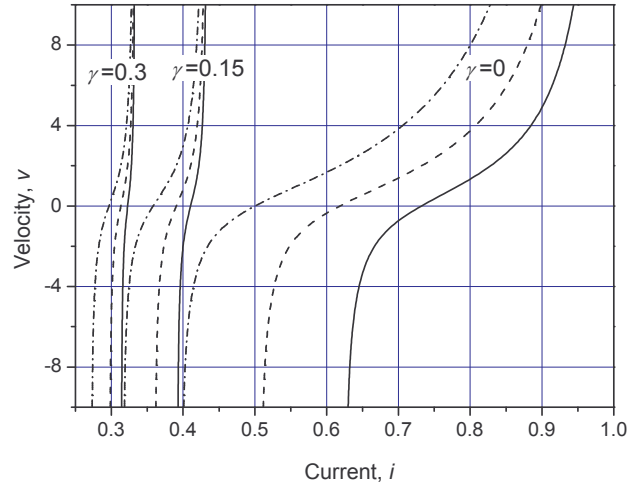

Fig. 2. Velocity of the normal zone propagation as a function of the current at various $\gamma$ and $\beta$: $\beta$ = 0.05 – solid line, $\beta$ = 0.1 – dashed line, $\beta$ = 0.2 – dash-dotted line; $\alpha$ = 20.

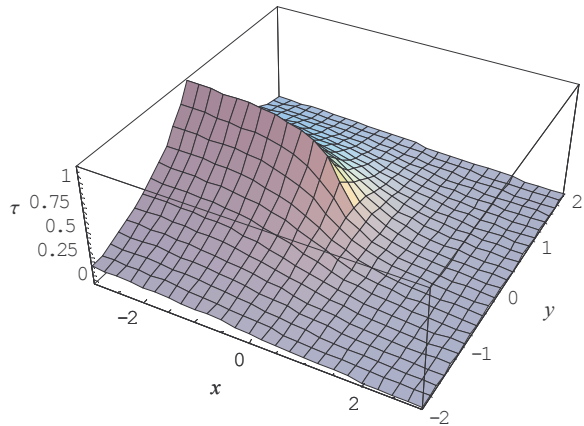

Fig. 3. Temperature distribution in the structure at $\alpha'_{ef}$ = 6 and $i_m$ = 0.434.

The dependence of the velocity $v$ on the current in films differs from that in wires (Fig. 4). In the case of a film the velocity is higher, especially, at a current above $i_m$.

### 3.2. Substrate of a finite width

The above expressions for the velocity of a normal zone have been obtained assuming that $l_s \gg 1$. As follows from (10), this approximation is correct if $l_s$ is larger than 3÷5. i.e. a half of the substrate width $L_s$ is larger than 3÷5 cm. Several prototypes of fault current limiters and switches were built using the substrates of smaller width [1,2,9,25].

Let us consider a straight infinite long film deposited in the center of the substrate of a finite width and assume that $h_2 \gg 1$. Due to symmetry of the task it is enough to consider a half of the substrate at $y > 0$, and the task is reduced to the solution of Eq. (9) with the boundary conditions

$$\left.\frac{\partial \tau_s}{\partial y}\right|_{y=0} = -\alpha'_{ef} i^2 \quad \text{and} \quad \left.\frac{\partial \tau_s}{\partial y}\right|_{y=l_s} = 0.$$

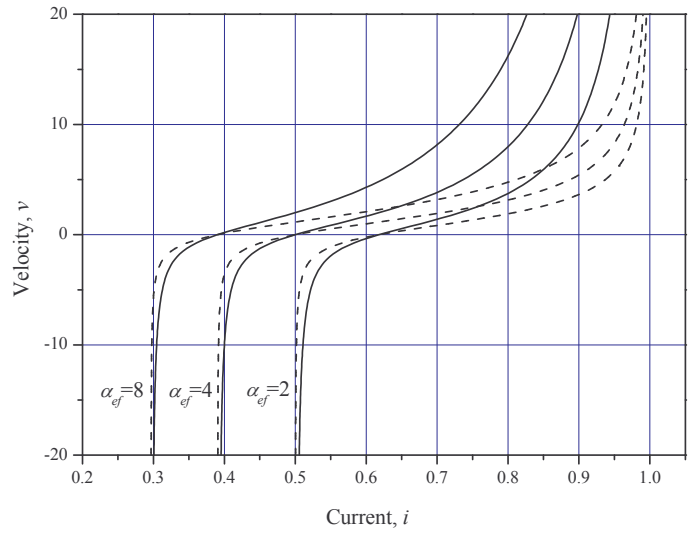

Fig. 4. Velocity of the normal zone propagation: solid line – velocity in a narrow film; dashed line – in a wire.

Using the solution of Eq. (12) presented in [24], we obtain the temperature distribution in the substrate:

$$\tau_s = \alpha'_{ef} i^2 \sum_{n=-\infty}^{\infty} FT(X, y - 2nl, v). \qquad (16)$$

Expression (16) differs from (13) and (14) by the terms with $n \neq 0$. Using the asymptotic for the modified Bessel function $K_0(x)$ at $x \to -\infty$, we can estimate that the sum of these terms decreases with $l_s$ rather than $\exp(-2l_s)$. This allows us to use the infinite approximation at $l_s > 2÷3$. The result is well confirmed by the numerical calculation based on Eq. (16) (Fig. 5). At $l_s = 2$ the values for the minimal current and propagation velocity differ from the results given by the approximation $l_s \to \infty$ only by a few percents.

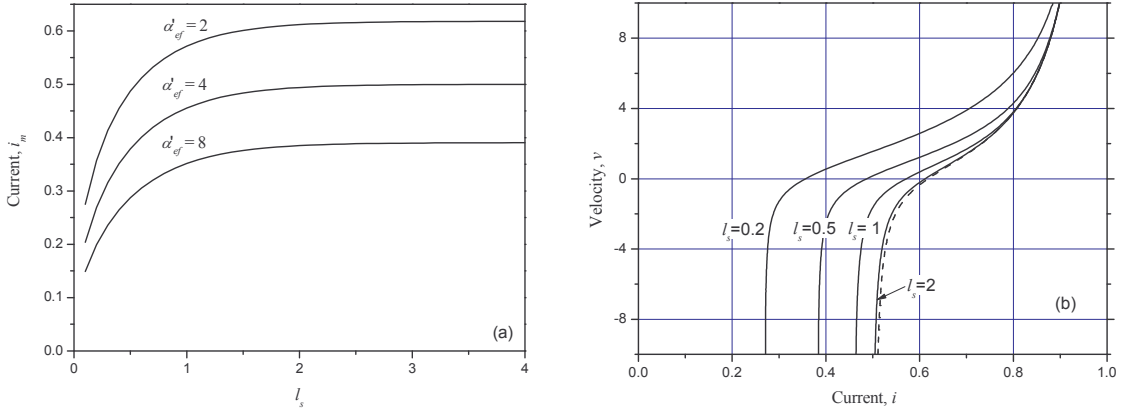

Fig. 5. Minimal current (a) and velocity (b) of the normal zone propagation in a film deposited on a finite width substrate. Dashed line in (b) presents the velocity at $l_s \to \infty$.

Note, that the minimal current is determined at $v = 0$ and, as follows from (16), for its estimation we can use the expression for a wire with an effective Stekly parameter depending on the substrate width. The decrease of the substrate width decreases the minimal current and increases the velocity. This is explained by the decrease of the cooled surface of the substrate, which increases the film temperature. The increase of the film temperature can be clearly illustrated by its dependence on the substrate width at $x \to -\infty$:

$$\tau_f = \alpha'_{ef} i^2 \coth(l_s).$$

With the increase, the velocity tends to the magnitude obtained at $l_s \to \infty$ (Fig. 5b) because, at a high velocity, the heat does not have time to propagate through the substrate. Fig. 6 presents the temperature distribution at the boundary between the normal and superconducting zones, i. e. at $X = 0$. Only a part of the substrate participates in the heat transfer to the coolant and in the heat propagation along the structure. The behavior of the wide structure is similar to that of a narrow substrate. The participation in cooling only of a substrate part can be an explanation of the increase of the difference between velocities in a film structure and a wire with the velocity (Fig. 4).

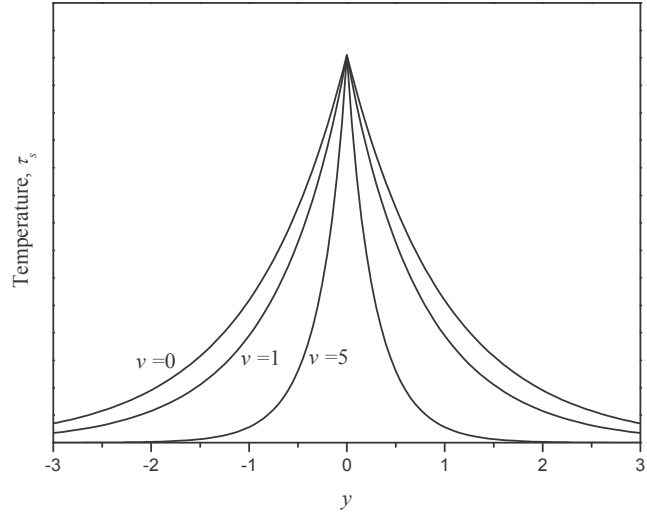

Fig. 6. Temperature as a function of $y$ at $X = 0$ and $l_s \to \infty$ for various velocities of the normal zone propagation. Here the temperatures at $X = 0$ and $y = 0$ are reduced to the same value.

*3.3. Normal zone propagation in a meander*

The application of superconducting films in switches, fault current limiters, self-limiting transformers and etc. requires a large resistance of the film in the normal state. To achieve larger resistance, the films of the spiral or meander topology are deposited on a large substrate [1,4,25,26]. Let us consider the normal zone propagation in a homogenous narrow film of the meander topology on a large substrate (Fig. 7). The normal zones can be nucleated near the current terminals due to losses in the contact of a normal wire and a film or at corners of the meander lines where the current density is very high [26]. Consider first the normal zone appearing near one of the current terminals and propagating to the second. We assume that all the distances $L_g$ between the long strips are the same and the strip length $L_f$ is large, e. i. $l_f = L_f/L_x \gg 1$ and $L_g/L_f \ll 1$. It means that the heat flux from the end short strips connecting the long strips can be neglected and the steady-state process of the normal zone propagation is described by a solution of Eq. (9). However, as distinct from the cases considered above, the normal zone is propagating in the film, which is previously heated due to losses in the strips that are already in the normal state. Considering that the strips are narrow, $l \ll 1$, the substrate temperature at any point is a sum of the temperatures given by the solutions of Eq. (9) separately for every strip (see Appendix). The propagation velocity is determined by the temperature at the boundary between the normal and superconducting zones, i. e. at $X = 0$ and $y = 0$, which is given by

$$\tau_f = \tau_s(0,0) = \alpha'_{ef} i^2 \left[ \frac{1 - e^{-l_g m}}{e^{l_g} - 1} + FT(0,0,v) \right], \quad (17)$$

where $m$ is the number of the strips, which are in the normal state. We use expressions (13) and (A6) for single films deposited on a large substrate at $h_2 \gg 1$.

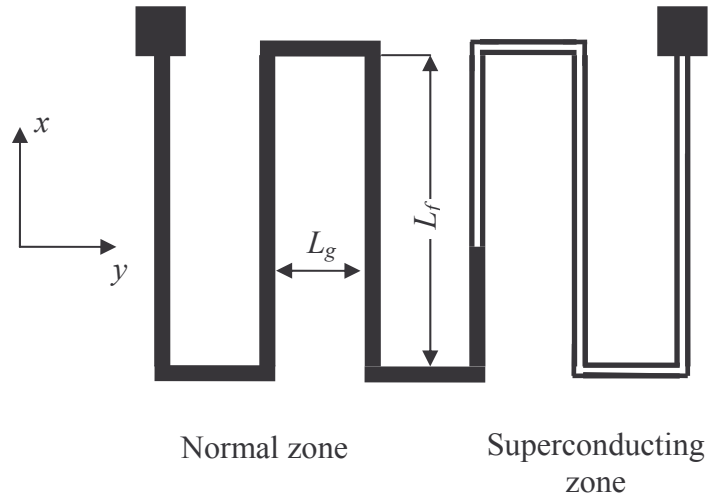

Fig. 7. Sketch of a film of the meander topology deposited on a wide substrate. Black squares note the current terminals.

The comparison of solutions (17) and (8) with account of (13) shows that the task of the normal zone propagation in the meander is reduced to the task for the case of a finite $h_2$ with $\alpha = \alpha'_{ef}$, $\beta = 1$ and $\gamma = \left(1 - e^{-l_g m}\right) / \left(e^{l_g} - 1\right)$ (Section 3.1). At $i =$const, the parameter $\gamma$ and, hence, the velocity increase with time if $v > 0$ and decrease in the opposite case, i.e. the velocity module increases with time in both cases.

Consider now the case where the normal zone is nucleated in one of the end stripes and propagates along two parallel long stripes (Fig. 8). If the task is symmetrical ($\alpha_1 = \alpha_2$, $k = 1$), the velocities in both strips are the same and determined by a solution of Eq. (11), where

$$\tau_f = \alpha'_{ef} \, i^2 \left[ FT(0,0,v) + FT(0, l_g, v) \right]. \tag{18}$$

Fig. 9 presents the velocity as a function of the distance between the strips. The velocity value lies between two asymptotic cases: $l_g = 0$ and a single film strip ($l_g \gg 1$). When the velocity increases, the curves are verging towards the curve obtained for a single film. At a high velocity, the heat does not have time to propagate through a substrate (Fig. 6) and the strips weakly influence one another.

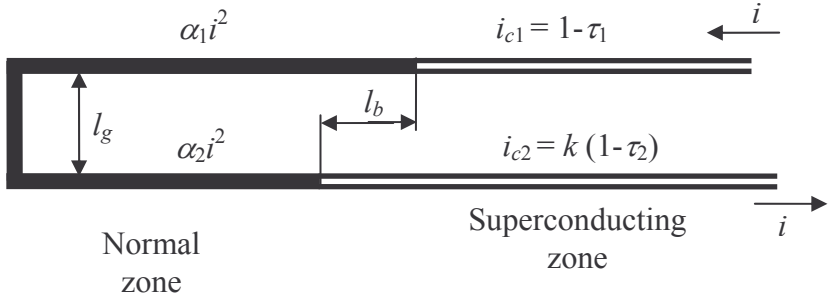

Fig. 8. Sketch of the normal zone propagation in a film of the meander topology deposited on a wide substrate. Here $\alpha_1$, $\alpha_2$, $i_{c1}$ and $i_{c2}$ are the effective Stekly parameters and critical currents for the first and second strips, respectively, $k$ is the ratio of these currents at $\tau = 0$.

Let us consider a meander where each stripe is homogeneous but their critical currents and losses are different. In this non-symmetrical case the velocities are different in various strips. The distance $l_b$ between the normal-superconducting zone boundaries increases with time. In the steady-state process, this distance is large, $l_b \gg 1$. The velocities of the normal zone are different but constant. The larger velocity is determined as for a single film, without taking into account of the second strip. The lower velocity can be estimated using expression (17) with $m = 1$.

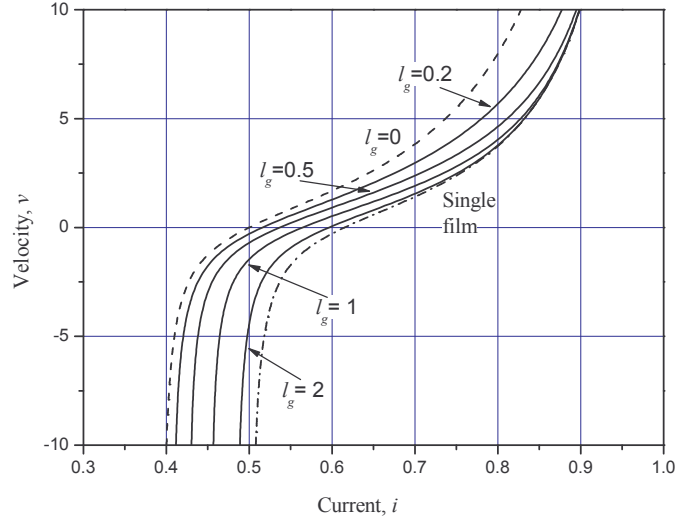

Fig. 9. Velocity of the normal zone propagation vs. current in a meander. Dashed line presents the velocity for $l_g = 0$, dash-dotted line – for $l_g \gg 1$.

Another possible regime is where the velocities of the normal zone propagation are the same in spite of the fact that the strips are different. The velocity $v$ and distance $l_b$ between the boundaries of the normal zones can be determined from the condition that the current $i$ equals to the critical value at the boundaries. The velocity and distance are given by a solution of the following system:

$$\alpha_1 i^2 FT(0,0,v) + \alpha_2 i^2 FT(l_b, l_g, v) = 1 - i, \tag{19}$$

$$\alpha_1 i^2 FT(-l_b, l_g, v) + \alpha_2 i^2 FT(0,0,v) = 1 - i/k. \tag{20}$$

The system gives a solution in a wide enough range of the values of $\alpha_1$, $\alpha_2$, $k$, and $i$. The boundaries of this range can be determined by the following way. One boundary is determined from the configuration presented in Fig. 8 at $l_b \gg 1$. In this case the system is simplified: the second term in the right hand side of (19) can be neglected and in Eq. (20) $FT(-l_b, l_g, v)$ can be replaced by $\exp(-l_g)$. The boundary relation between the parameters is

$$\alpha_1 i^2 e^{-l_g} + \frac{\alpha_2(1-i)}{\alpha_1} = 1 - i/k . \tag{21}$$

Another boundary is determined from the condition that the normal zone in the second strip is forward:

$$\alpha_2 i^2 e^{-l_g} + \frac{\alpha_1(k-i)}{\alpha_2 k} = 1 - i . \tag{22}$$

Note that the steady-state process of the normal zone propagation along two parallel strips is stable. For example, the distance $l_b$ increased (Fig. 8). This leads to a decrease of the contribution of the second strip in the temperature $\tau_1$ which causes lowering the velocity in the first strip. On the other hand the contribution of the first strip in the temperature $\tau_2$ increases with $l_b$. Hence, rise of $l_b$ causes an increase of the normal zone velocity in the second strip. Similar reasoning shows that a decrease of $l_b$ leads to an increase of the velocity in the first strip and to a decrease in the second one. Thus, fluctuations in the distance $l_b$ decrease with time.

The results obtained for two parallel strips can be used for the analysis of the normal zone propagation in the multifilament coated superconductors where electrical contact between superconducting strips and a conducting substrate is absent. To estimate the velocities of the normal zone propagation, the expressions obtained above have to be slightly changed with taking into account the distinction of currents in various filaments. The additional approximation is related to the assumption that the currents in the filaments are constant. This assumption is valid if the normal zones are sufficiently long.

## 4. Films deposited on a conducting substrate

If there is the electrical contact between a film and a conducting substrate, the substrate shunts the film in the normal state, and a part of the current flows through the substrate. This leads to a decrease of losses in the film and, on the other hand, produces the Joule losses in the substrate. Account of the losses in the substrate leads to appearance of the additional terms in Eq. (12), which is now represented as

$$\frac{\partial^2 w}{\partial X^2} + \frac{\partial^2 w}{\partial y^2} = \left(1 + \frac{v^2}{4}\right) w - q \ , \tag{23}$$

where $q = \dfrac{\rho_s \Delta_s J_s^2}{H_z(T_c - T_0)}$ is the normalized loss density in the substrate, $J_s$ is the current density in the substrate and $\rho_s$ is its resistance.

Using the solution for Eq. (23) presented in [24], the temperature of the substrate with the width $2l_s$ at $h_2 \gg 1$ can be presented in the following form:

$$\tau_s(X,y) = \alpha'_{ef} \int_{-\infty}^{0} i_f^2 e^{\frac{v(\xi - X)}{2}} G(X,y,\xi,0) d\xi + \int_{0}^{l_s} \int_{-\infty}^{\infty} q(\xi,\eta) e^{\frac{v(\xi - X)}{2}} G(X,y,\xi,\eta) d\xi d\eta , \tag{24}$$

where

$$G(X,y,\xi,\eta) = \frac{1}{2\pi} \sum_{n=-\infty}^{\infty} [K_0(s\delta_{n1}) + K_0(s\delta_{n2})],$$

$$s = \sqrt{1 + \frac{v^2}{4}}, \quad \delta_{n1,2} = \sqrt{(X-\xi)^2 + (y - 2nl_s \pm \eta)^2} \ ,$$

$i_f$ is the film current which is a function of $X$ even for the stepwise voltage-current characteristic (7) of the film. The determination of the current distribution in the substrate and of the film current is an intricate problem, which can be solved numerically. Here we consider the simplest case where $J_s$ and the film current $i_f$ are stepwise functions. At $X < 0$ they are constants, and at $X > 0$ $J_s = 0$ and $i_f = i$. The normal zone velocity can be determined from the condition that at $X = 0$ the current $i$ is equaled to the critical value $1-\tau_s(0,0)$. This picture of the current distribution is correct only if the substrate and film widths are equal. The accuracy of this model decreases with increase of the substrate width. Taking into account that the structure temperature is determined by the integrals of $J_s$ and $i_f$, we expect that for a sufficiently narrow substrate the error given by the simplest model would be acceptable. The temperature at $X = 0$ and $y = 0$ is

$$\tau_s(0,0) = \frac{i^2}{(1+r)^2} \int_{-\infty}^{0} e^{\frac{v\xi}{2}} \left[ r^2 \alpha'_{ef} G(0,0,\xi,0) + \alpha_s \int_0^{l_s} G(0,0,\xi,\eta) d\eta \right] d\xi, \qquad (25)$$

where $r = R_s/R_f$ is the ratio of the substrate $R_s$ and film $R_f$ resistances, $\alpha_s = \dfrac{I_c(T_0)^2 R_s}{2 l_s H_z (T_c - T_0)}$.

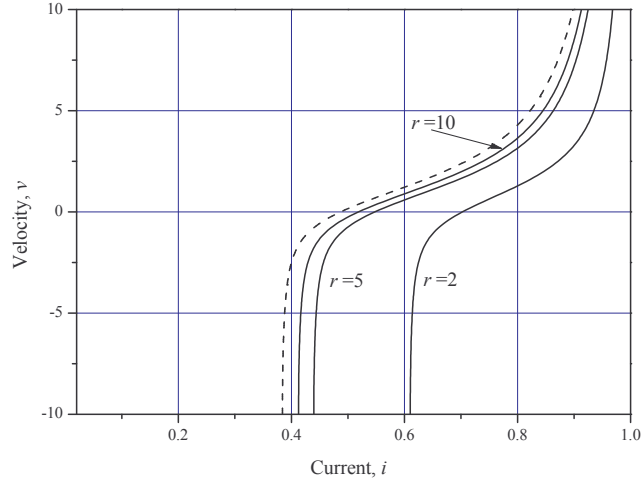

Fig. 10. Velocity of the normal zone vs. current in the structure containing a superconducting film deposited on a conducting substrate at $\alpha'_{ef} = 2$, $l_s = 0.5$, $\alpha_s = 0.5$. Dashed line presents the velocity vs. a current for a film deposited on a dielectric substrate.

A conducting substrate shunts a superconducting film in the normal state that is equivalent to the decrease of parameter $\alpha'_{ef}$ (see Figs. 4 and 10). The conducting substrate with even relatively high resistance markedly decreases the velocity and increases the minimal current of the normal zone propagation (Fig. 10, lines for $r=10$ and $r=5$). The effect decreases with the increase of the velocity. This is explained by a finite rate of the heat propagation through the substrate (see section 3.2).

5. Conclusion

The proposed mathematical model allowed us to obtain analytical expressions using which the influence of various factors on the velocity of the normal zone propagation were analyzed. These expressions were used to estimate the velocity dependence on thermal contact resistance between a film and a substrate, width and conductivity of the substrate, film topology. It was shown that

- to estimate the minimal current of the normal zone propagation, the well-known expression for a wire can be used with introducing an effective Stekly depending on the thermal contact between a film and a substrate, the substrate width, and the film topology. However, the expression for the normal zone velocity for a wire gives unacceptable errors for a film especially for high velocity values.
- the dependence of the velocity on the substrate width decreases with the velocity increase and at $v > 10$ the dependence practically vanishes. This can be explained by a finite rate of the heat propagation through a substrate.
- in a meander, the velocity is increasing with time due to a heating of the structure by losses in a part of the meander which is in the normal state.
- in parallel films, the normal zones can propagate with the same velocity even when the properties (critical currents, the Stekly parameters) of the films are different.
- if a film is deposited on a conducting substrate, the influence of shunting on the velocity is pronounced even at a relatively high substrate resistance.

**Appendix**

At $h_2 \gg 1$ and $l \ll 1$ the film cooling is realized only by the heat fluxes through the substrate. These fluxes are determined by the temperature derivatives and are independent of the temperature. A temperature in any point of the substrate can be presented as a sum of the temperatures each one of them is determined separately for every strip. Let us illustrate this using the simplest case. Two infinite long narrow films with uniform loss densities $w_1$ and $w_2$ are deposited on a large substrate at $y = l_1$ and $y = -l_1$, respectively. In the stationary regime the temperature distribution in the structure is determined by the equation

$$\frac{d^2\tau_s}{dy^2} = -\tau_s,  \tag{A1}$$

with the boundary conditions given by condition of the temperature continuity at any point and

$$\left.\frac{d\tau_s}{dy}\right|_{y\to-\infty} = 0, \quad \left.\frac{d\tau_s}{dy}\right|_{y\to\infty} = 0,$$

$$\left.\frac{d\tau_s}{dy}\right|_{y\to-l_1-0} - \left.\frac{d\tau_s}{dy}\right|_{y\to-l_1+0} = w_2, \tag{A2}$$

$$\left.\frac{d\tau_s}{dy}\right|_{y\to l_1-0} - \left.\frac{d\tau_s}{dy}\right|_{y\to l_1+0} = w_1. \tag{A3}$$

The solution of Eq. (A1) is

$$\tau_s = \begin{cases} Ae^y, & y < -l_1, \\ A_1 e^y + B_1 e^{-y}, & -l_1 < y < l_1, \\ Be^{-y}, & y > l_1, \end{cases} \tag{A4}$$

where $A$, $A_1$, $B$, and $B_1$ are the integration constants.

The solution is

$$\tau_s = \frac{1}{2}\begin{cases} w_1 e^{-l_1+y} + w_2 e^{l_1+y}, & y < -l_1, \\ w_1 e^{-l_1+y} + w_2 e^{-l_1-y}, & -l_1 < y < l_1, \\ w_1 e^{l_1-y} + w_2 e^{-l_1-y}, & y > l_1. \end{cases} \tag{A5}$$

The solutions for each single film are, respectively:

$$\tau_1 = \frac{w_1}{2}\begin{cases} e^{-l_1+y}, & y < l_1, \\ e^{l_1-y}, & y > l_1, \end{cases} \text{ and } \tau_2 = \frac{w_2}{2}\begin{cases} e^{l_1+y}, & y < -l_1, \\ e^{-l_1-y}, & y > l_1. \end{cases} \tag{A6}$$

The comparison of expressions (A5) and (A6) gives $\tau_s = \tau_1 + \tau_2$ at any point of the structure.